\documentclass{article}

\usepackage{microtype}
\usepackage{graphicx}
\usepackage{subfigure}
\usepackage{booktabs} 

\usepackage{hyperref}

\usepackage[accepted]{icml2020}

\icmltitlerunning{TRL4ML}

\begin{document}

\twocolumn[
\icmltitle{Technology Readiness Levels for AI \& ML}



\icmlsetsymbol{equal}{*}

\begin{icmlauthorlist}
\icmlauthor{Alexander Lavin}{aug}
\icmlauthor{Gregory Renard}{aug}
\end{icmlauthorlist}

\icmlaffiliation{aug}{Augustus Intelligence}

\icmlcorrespondingauthor{Alexander Lavin}{alexander.lavin@augustusai.com}

\icmlkeywords{Machine Learning, Systems Engineering}

\vskip 0.3in
]



\printAffiliationsAndNotice{}  

\begin{abstract}
The development and deployment of machine learning systems can be executed easily with modern tools, but the process is typically rushed and means-to-an-end. The lack of diligence can lead to technical debt, scope creep and misaligned objectives, model misuse and failures, and expensive consequences. Engineering systems, on the other hand, follow well-defined processes and testing standards to streamline development for high-quality, reliable results. 
The extreme is spacecraft systems, where mission critical measures and robustness are ingrained in the development process.
Drawing on experience in both spacecraft engineering and AI/ML (from research through product), we propose a proven systems engineering approach for machine learning development and deployment. Our \textit{Technology Readiness Levels for ML (TRL4ML)} framework defines a principled process to ensure robust systems while being streamlined for ML research and product, including key distinctions from traditional software engineering.
Even more, TRL4ML defines a common language for people across the organization to work collaboratively on ML technologies.
\end{abstract}

\section{Introduction}

The development of machine learning (ML) models is often narrow- and near-sighted, only considering the publication target or minimum viable product requirements. 
A main concern is models are typically trained and tested on only a handful of curated datasets, without measures and safeguards for future scenarios. 
Code quality is typically subpar and poorly documented, a technical debt that is exacerbated because future users are rarely the original researchers or developers. Models and algorithms for deployment are integrated in a software stack that is robust and documented, but without regard for the inherent stochasticity\footnote{Consider the massive effect random seeds have on deep reinforcement learning model performance, shown by \citet{Henderson2018DeepRL}} and failure modes of the hidden ML components.

Other domains of engineering, such as civil and spacecraft, follow well-defined processes and testing standards to streamline development for high-quality, reliable results. \textit{Technology Readiness Level (TRL)} is a systems engineering protocol for deep tech and scientific endeavors at scale, ideal for integrating many interdependent components \textit{and} cross-functional teams of people. No surprise TRL is standard process and parlance in NASA and DARPA \cite{Nasa2003NASASE}.

For a spaceflight project there are several defined phases, from pre-concept to prototyping to deployed operations, each with a series of development cycles and reviews. This is in stark contrast to machine learning and software workflows, which promote quick iteration, rapid deployment, and simple linear progressions. Yet the NASA technology readiness process is overkill. We aim to bring systems engineering to machine learning by defining and putting into action a lean \textit{Technology Readiness Levels for ML (TRL4ML)} framework. We draw on decades of AI development, from research through production, across domains and applications: for example, computer vision in medical diagnostics and factory robotics, NLP in commerce and social media, streaming time-series in predictive maintenance and finance.

In this paper we define our proven framework for developing and deploying robust ML systems, with a real example of advancing a novel algorithm from R\&D through productization and deployment within a massive system.
Our aim is to standardize TRL4ML to enable ML and SWE teams to develop principled, robust AI technologies.
Ultimately, TRL4ML gets people across the organization speaking the same language.


\section{TRL4ML}
\textit{TRL4ML} defines technology readiness levels (TRLs) \cite{Nasa2003NASASE} to guide and communicate machine learning development and deployment. A TRL represents the maturity of a model or algorithm\footnote{Note we use ``model" and ``algorithm" somewhat interchangeably when referring to the technology under development. The same TRL4ML process applies for e.g. a machine translation model or an algorithm for A/B testing.}, data pipes, software module, or composition thereof; a typical ML system consists of many interconnected subsystems and components, and the TRL of the systems is the lowest level of its constituent parts.
The levels are briefly defined as follows, and elucidated with an example project in Fig. \ref{fig:bo_example}:

\begin{figure*}[!h]
  \centering
  {\includegraphics[width=0.9\linewidth]{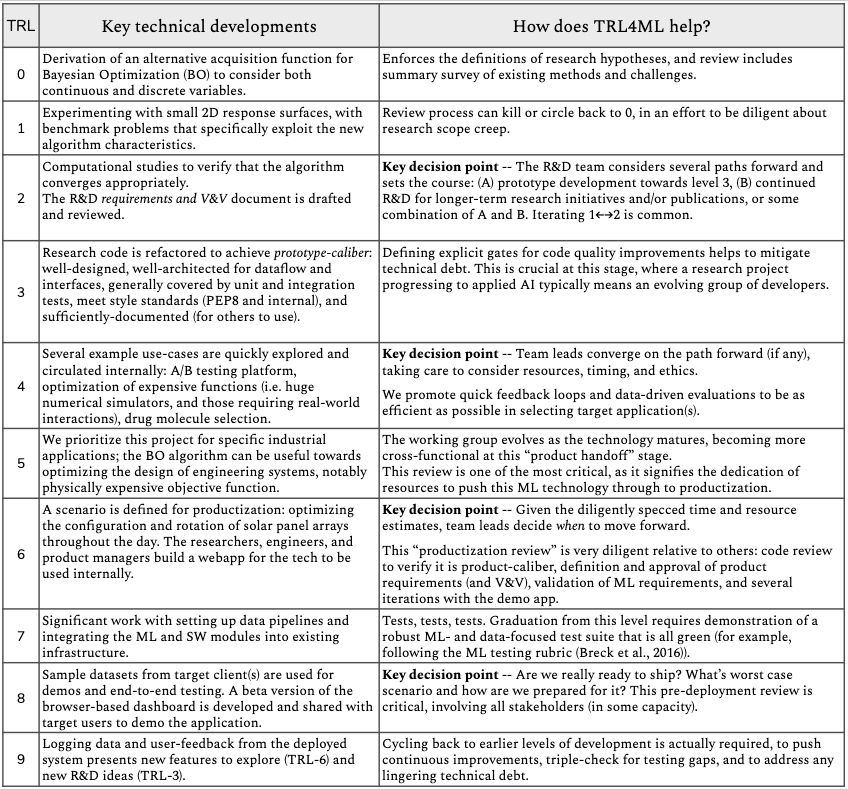}}
  \caption{Highlights from a real-world example utilizing TRL4ML to develop a novel Bayesian Optimization \cite{Shahriari2016BOReview} algorithm from low-level research through prototyping, productization, and deployment.}
  \label{fig:bo_example}
\end{figure*}

\vspace*{-\baselineskip}
\vspace{2mm}
\paragraph{Level 0 - Brainstorming} \textit{A stage for greenfield research.}

\vspace*{-\baselineskip}
\vspace{3mm}
The outcome is a set of concrete ideas with sound maths, to pursue through low-level experimentation in the next stage. To graduate, the basic principles, hypotheses, and research plans need to be stated, referencing relevant papers. The reviewer here is solely the research team lead.

\vspace*{-\baselineskip}
\vspace{2mm}
\paragraph{Level 1 - Goal-Oriented Research} \textit{Moving from basic principles to practical use.}

\vspace*{-\baselineskip}
\vspace{3mm}
Here we design and run low-level experiments to analyze model/algorithm properties, which need to pass a peer-review process before graduating to level 2 -- the review panel includes additional members of the research team.

\vspace*{-\baselineskip}
\vspace{2mm}
\paragraph{Level 2 - Proof of Principle (PoP) Development} \textit{Active R\&D is initiated.}

\vspace*{-\baselineskip}
\vspace{3mm}
The models run in testbeds: simulated environments and/or surrogate data that closely matches the conditions and data of real scenarios -- note these are not product-driven. An important deliverable at this stage is the formal requirements document (with well-specified verification and validation steps).
The culmination of this stage is often a bifurcation: some work moves to applied AI, while some circles back for more research.

\vspace*{-\baselineskip}
\vspace{2mm}
\paragraph{Level 3 - System Development} \textit{Sound software engineering.}

\vspace*{-\baselineskip}
\vspace{3mm}
Here we have checkpoints that push code development towards interoperability, reliability, maintainability, extensibility, and scalability. In TRL4ML we develop with the mindset that research code will be thrown away when the project development calls for more legitimate software engineering.
The level 3 review includes teammates whom focus more on applied AI and engineering.

\vspace*{-\baselineskip}
\vspace{2mm}
\paragraph{Level 4 - Proof of Concept (PoC) Development} \textit{Demonstration in a real scenario.}

\vspace*{-\baselineskip}
\vspace{3mm}
This stage is the seed of application-driven development; for many organizations this is the first touch-point with product managers and stakeholders beyond the R\&D group.
In review, we demonstrate the utility towards one or more practical applications, taking care to communicate assumptions and limitations.
Ideally the organization has an AI ethics review process, which would be appropriate at this stage (as the AI capabilities and datasets are known).

\vspace*{-2mm}
\paragraph{Level 5 - Machine Learning ``Capability"} \textit{The R\&D to product handoff.}

An interdisciplinary working group is defined, as we start developing the tech in the context of a larger real-world process – i.e., transitioning the model or algorithm from an isolated solution to a module of a larger application.
Graduation from level 5 should be difficult, as it signifies the dedication of resources to push this ML technology through to productization.

\vspace*{-2mm}
\paragraph{Level 6 - Application Development} \textit{Robustification of ML modules, specifically towards one or more use-cases.}

The main work here is significant software engineering to bring the code up to \textit{product-caliber}, as well as defining product-specific requirements and data pipelines spec. 

\vspace*{-3mm}
\paragraph{Level 7 - Integrations} \textit{ML infrastructure, product platform, data pipes, security protocols.}

For integrating the technology into existing production systems, we recommend the working group has a balance of infrastructure engineers \textit{and} applied AI engineers -- we find this stage of development is vulnerable to latent model assumptions and failure modes.
The review should focus on the data pipelines and test suites; a scorecard like the ML Testing Rubric is useful \cite{Breck2016MLScore}. We stress the need for tests that run use-case specific critical scenarios and data-slices -- a proper risk-quantification table will highlight these.

\vspace*{-4mm}
\paragraph{Level 8 - Flight-ready} \textit{The end of system development.}

The technology is demonstrated to work in its final form and under expected conditions. There should be additional tests implemented at this stage covering deployment aspects: A/B tests, blue/green deployment tests, shadow testing, canary testing, and others. 
Review panel is representative of the full slate of stakeholders.
We diligently walk through every technical and product requirement, and corresponding validations.

\vspace*{-4mm}
\paragraph{Level 9 - Deployment} \textit{Monitoring the current version, improving the next.}

Maintenance engineering (i.e. monitoring and update methods) takeover; CI/CD should regularly stress test the system, and regression tests on ML components send logs to relevant applied \textit{and} research engineers. 
There is a defined communication path for user feedback, without roadblocks to R\&D; we encourage real-world feedback all the way to research, providing valuable problem constraints and perspectives.

\begin{figure*}[!th]
  \centering
  {\includegraphics[width=0.95\linewidth]{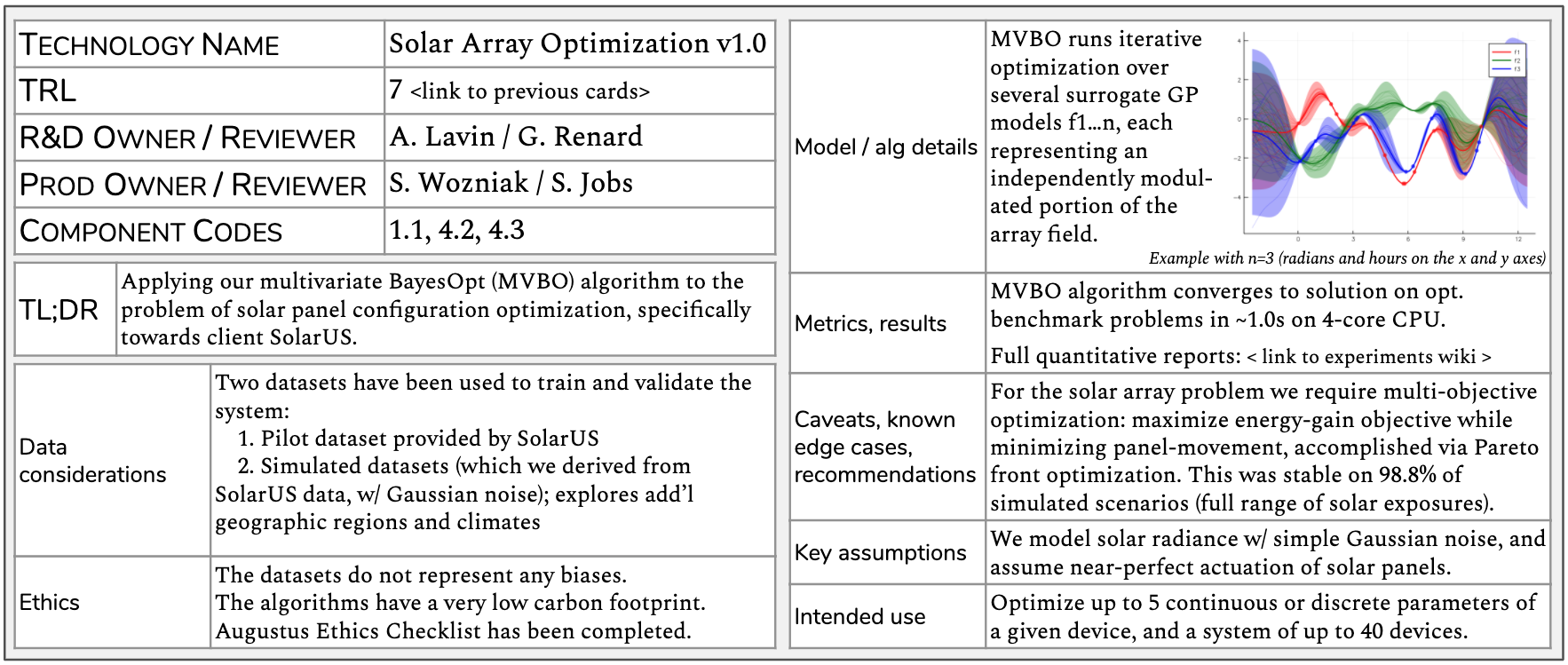}}
  \caption{The maturity of each model or algorithm is tracked via \textit{TRL cards}. Here is an example, reflecting a Bayesian Optimization algorithm for industrial process optimization; note this is a subset of the full card.}
  \label{fig:trl_card}
\end{figure*}

\subsection{Key components in the process}

\paragraph{Level reviews}
At the end of each stage is a dedicated review period: present the tech developments and their validations, make key decisions on path(s) forward (or backward), and debrief the process.\footnote{TRL4ML should include regular debriefs and meta-evaluations such that process improvements can be made in a data-driven, efficient way (rather than an annual meta-review). TRL4ML is a high-level framework that each organization should operationalize in a way that suits their specific capabilities and resources.} 
The designated reviewers will ``graduate" the technology to the next level, or provide a list of specific tasks that are still needed (ideally with quantitative remarks).
After graduation at each level, the working group does a brief post-mortem; we find that a quick day or two pays dividends in cutting away technical debt and improving team processes.

\vspace*{-\baselineskip}
\vspace{2mm}
\paragraph{TRL cards}
In Fig. \ref{fig:trl_card} we succinctly showcase a key deliverable: \textit{TRL cards}.
The model cards proposed by Google \cite{Mitchell2019ModelCF} are a useful development for external user-readiness with ML.
On the other hand, our TRL cards are more like ``report cards" that grow and improve upon graduating levels, and provide a means of inter-team and cross-functional communication.
The content of a TRL card is roughly in two categories: project info, and implicit knowledge. 
The former clearly states info such as project owners and reviewers, development status, and semantic versioning (for code, models, and data). In the latter category are specific insights that are typically siloed in the ML development team but should be communicated to other stakeholders: modeling assumptions, dataset biases, corner cases, etc.


\vspace*{-\baselineskip}
\vspace{2mm}
\paragraph{Risk mitigation}
Identifying and addressing risks in a software project is not a new practice. However, akin to the TRL4ML roots in spacecraft engineering, risk is a ``first-class citizen" here. In the definition of technical and product requirements, each entry has a calculation of the form $risk = p(failure) \times value$, where the value of a component is an integer $1-10$. Being diligent about quantifying risks across the technical requirements is a useful mechanism for flagging ML-related vulnerabilities that can sometime be hidden by layers of other software. 
TRL4ML also specifies that risk quantification and testing strategies are required for sim-to-real development. That is, there is nearly always a non-trivial gap in transferring a model or algorithm from a simulation testbed to the real world. Requiring explicit sim-to-real testing steps in the workflow helps mitigate unforeseen (and often hazardous) failures.

\vspace*{-\baselineskip}
\vspace{2mm}
\paragraph{Non-linear, non-monotonic paths} 
We observe many projects benefit from cyclic paths, dialing components of a technology back to a lower level following the stage review. 
Our framework not only uses cycles, but actively discourages the straight path approach that is typically assumed in ML projects.
It's also important to note that most projects do not start at level 0; very few ML companies engage in this low-level theoretical research. For example, a team looking to use an off-the-shelf object recognition model would start that technology at level 4. However no technology can skip levels after the TRL4ML process has been initiated.

\vspace*{-\baselineskip}
\vspace{2mm}
\paragraph{Quantifiable progress}
By defining technology maturity in a quantitative way, TRL4ML enables teams to accurately and consistently define their ML progress metrics; OKRs and KPIs can be defined as achieving certain levels in a given period of time.
Even more, meta-review of TRL4ML progress over multiple projects can provide useful insights at the organization level. For example, analysis of the time-per-level and the most frequent development paths/cycles can bring to light operational bottlenecks.
Compared to conventional SWE metrics based on sprint stories and tickets, or time-tracking tools, TRL4ML provides a more accurate analysis of ML workflows.

\vspace*{-\baselineskip}
\vspace{2mm}
\section{Discussion}

There are several key areas where machine learning (ML) development is unique from software engineering (SWE). 
For instance, the behavior of ML systems is learned from data, not specified directly in code. The data requirements around ML (i.e., data discovery, management, and monitoring) adds significant complexity not seen in other types of SWE.
Not to mention an array of ML-specific failure modes; for example, models that become mis-calibrated due to subtle data distributional shifts in the deployment setting, resulting in models that are more confident in predictions than they should be.
These are a couple instances of broader themes we've observed, where ML systems depart from the rest of SWE. 
A recent case study from Microsoft Research \cite{Amershi2019SoftwareEF} similarly identifies a few themes. Also related to our work, Google teams have proposed ML testing recommendations \cite{Breck2016MLScore} and validating the data fed into ML systems \citet{Breck2019DataVF}.
These analyses provide useful insights, but they do not provide a holistic, regimented process for the full ML lifecycle.

We've introduced \textit{TRL4ML}, a proven systems engineering process for machine learning. 
Our hope is the framework is adopted broadly in AI/ML organizations, and that ``technology readiness levels" becomes common nomenclature across stakeholders -- from researchers and engineers to sales-people and CEOs.


\bibliography{trl_icml}
\bibliographystyle{icml2020}

\end{document}